# Introduction to Set Shaping Theory

Solomon Kozlov[1]

**Abstract:** given a source defined by an ensemble $X = (x; A; P)$, we call $X^N$ the set that contains all the possible strings $x = \{x_1, \ldots, x_N\}$ generated by $X$. The Set Shaping Theory studies the bijection functions $f: X^N \to Y^{N+k}$ With $K$ and $N \in \mathbb{N}^+$, $|X^N| = |Y^{N+k}|$ and $Y^{N+K} \subset X^{N+K}$. In this article, we analyze the functions $f$ in which the set $Y^{N+K}$ contains the strings with the least information content belonging to the set $X^{N+K}$. The results obtained show how this type of function can be useful in data compression.

## Introduction

In this article, we introduce the Set Shaping Theroy whose objective is the study of the bijection functions that transform a set $X^N$ of strings of length $N$ into a set $Y^{N+K}$ of strings of length $N+K$ with K and $N \in \mathbb{N}^+$, $|X^N| = |Y^{N+k}|$ and $Y^{N+K} \subset X^{N+K}$. In particular, we will analyze the functions in which the set $Y^{N+K}$ contains the strings with less information content belonging to the set $X^{N+K}$. The analysis of the results shows how this type of function can be useful in data compression.

## Methods

In this article, we use the concepts and functions developed by C.E.Shannon [1] that represent the basis of information theory. Given a source defined by an ensemble $X = (x; A; P)$, where $x$ is the value of the random variable, $A = \{a_1, a_2, \ldots a_I\}$ are the possible values of $x$ (states) and $P = \{p_1, p_2 \ldots p_I\}$ is the probability distribution of the states $P(a_i) = p_i$ with $\sum_{i=1}^{I} pi = 1$.

The entropy of $X$, denoted $H$, is defined as:

$$H(X) = -\sum_i p_i \, log_b p_i$$

We call $X^N$ the set that contains all possible strings $x = \{x_1, \ldots, x_j, \ldots, x_N\}$ generated by $X$.

**Definition 1:** *We call f the bijection function on the set $X^N$ defined as*:

$$f: X^N \to Y^{N+k}$$

$K, N \in \mathbb{N}^+$, $|X^N| = |Y^{N+k}|$ and $Y^{N+K} \subset X^{N+K}$.

$f(x) = y$ with $x = \{x_1, \ldots, x_N\}$ and $y = \{y_1, \ldots, y_{N+K}\}$, $x \in X^N$ and $y \in Y^{N+K}$

The function $f$ defines from the set $X^{N+K}$ a subset of size equal to $|X|^N$. This operation is called "Shaping of the source" because what is done is to make null the probability of generating some sequences belonging to the set $X^{N+K}$.

**Definition 2:** *The parameter K is called the shaping order of the source and represents the difference in length between the sequences belonging to $X^N$ and the transformed sequences belonging to $Y^{N+k}$.*

Given a source $X = (x; A; P)$, and a string $x_i = \{x_1, \ldots, x_j, \ldots, x_N\}$, we define its information content:

$$I(x_i) = -\sum_{j=1}^{N} \log_2 p(x_j)$$

[1] Author correspondence: solomon.kozlov@mailfence.com

The probability $P(x_i)$ that the source $X$ generates the sequence $x_i$ is:

$$P(x_i) = \prod_{j=1}^{N} p(x_j)$$

**Definition 3:** *we call the average information content of a sequence generated by a source $X = (x; A; P)$ the summation of the product between the information content of the sequences belonging to $X^N$ is their probability:*

$$I(x) = \sum_{i=1}^{|X|^N} P(x_i) I(x_i) \quad (1)$$

**Remark 1:** As $N$ tends to infinity $I(x)$ tends to *NH(X)*. Indeed, when $N$ becomes large the contribution to the value of the function (1) derives almost exclusively from the strings belonging to the typical set [2]. With typical set we mean, the set of strings whose information content is close to *NH(X)*.

This function is essential to understand the advantages of applying $f$, because this function transforms the strings $x \in X^N$ into the strings $y \in Y^{N+K}$ consequently, the average information content changes as follows:

$$I(y) = \sum_{i=1}^{|X|^N} P(x_i) I(y_i) \quad (2)$$

Where $P(x_i)$ remains unchanged but the information content of the string changes.

**Definition 4:** *We call $f_m$ the bijection function on the set $X^N$ defined as*:

$$f_m : X^N \to Y^{N+k}$$

$K, N \in \mathbb{N}^+$, $|X^N| = |Y^{N+k}|$, $Y^{N+K} \subset X^{N+K}$, $X^{N+K} - Y^{N+K} = C^{N+K}$, $\forall y \in Y^{N+K}$ and $\forall c \in C^{N+K}$ $I(y) < I(c)$ and $I(y_i) < I(y_{i+1})$ $\forall y \in Y^{N+K}$.

**Remark 2:** The function $f_m$ transforms the set $X^N$ into the set $Y^{N+k}$ composed of $|X^N|$ strings with less information content belonging to $X^{N+K}$. Consequently, each string belonging to the complementary set of $Y^{N+K}$ has a greater information content than any string belonging to $Y^{N+k}$.

Wanting to apply this type of function to problems concerning data compression, the $f_m$ functions are the most interesting to be analyzed.

**Results**

Given a source defined by an ensemble $X = (x; A; P)$ with a uniform probability distribution, we will apply the function $f_m$ to the set $X^N$, which contains all possible strings of length $N$ produced by $X$, and compare the values of $I(x)$ and $I(y)$ with $f_m(x) = y$

We start by analyzing strings of lengths equal to $= |A|$, $K=1$ and $|A|$ variable between 2 and 7. Consequently, the length of the strings $x \in X^{|A|}$ is $|A|$ and having chosen $K=1$ the strings $y \in X^{|A|+1}$ have length $|A| + 1$. Therefore, for example, if $|A| = 2$ the strings $x \in X^2$ have length 2 instead the strings $y \in Y^3$ have length 3. Being $N$ very small it is possible to calculate the value of $I(x)$ and $I(y)$ exactly. The first column shows the cardinality of $A$. The second column shows the value of $I(x)$, the third shows the value of $I(y)$ and finally the fourth shows the difference $I(x)$-$I(y)$.

| $|A|$ | $I(x)$ $N = |A|$ | $I(y)$ $N = |A| + 1$ | $I(x)$-$I(y)$ |
|---|---|---|---|
| 2 | 1,000 | 1,377 | -0,377 |
| 3 | 2,893 | 2,885 | 0,009 |
| 4 | 5,296 | 5,050 | 0,246 |
| 5 | 8,070 | 7,708 | 0,362 |
| 6 | 11,137 | 10,223 | 0,915 |
| 7 | 14,448 | 13,387 | 1,061 |

*Table 1: The average information content I(x) and I(y) in bits calculated for different values of $N = |A|$ and K = 1.*

Having chosen such short string lengths, we have a value of $I(x)$ that differs greatly from *NH(X)*. This result is normal since for these values of $N$ the value calculated with the formula (1) depends very much on strings with information content less than *NH(X)*. Observing the data in table 1, we notice an unexpected result, indeed for values of $|A| > 2$ the average information content *I(y)* is less than *I(x)*.

Now, let's increase the length of the strings to 100 and keep the value of *K* at 1. Thus, the strings $x \in X^{100}$ have length 100, consequently having chosen *K=1* the strings $y \in X^{101}$ have length 101. In this case, given the length of the strings, the exact calculation of *I(x)* and *I(y)* is very complex, so we estimate these values using the Mote Carlo method [3] and [4]. The data reported in table 2 concern the simulation of 1000000 of strings of length 100 generated by a source $X = (x; A; P)$ with a uniform probability distribution and *|A|* variable between 2 and 10. The first column shows the cardinality of *A*. The second column shows the value of *I(x)*, the third shows the value of *I(y)* and finally the fourth shows the difference *I(x)-I(y)*.

Analyzing the data in table 2, we note that for this value of *N* the value *I(x)* approximates *NH(X)*. Indeed, as mentioned, increasing the length of the strings the contribution to formula (1) almost exclusively depends on the strings with information content close to *NH(X)*. Also in this case for *|A| > 2* the average information content *I(y)* is less than *I(x)*. Hence, this result does not depend on the length of the strings but also remains by increasing *N*.

| \|A\| | I(x) N=100 | I(y) N=101 | I(x)-I(y) |
|---|---|---|---|
| 2 | 99,275 | 99,660 | -0,385 |
| 3 | 157,044 | 157,034 | 0,011 |
| 4 | 197,816 | 197,331 | 0,485 |
| 5 | 229,279 | 228,315 | 0,964 |
| 6 | 254,850 | 253,436 | 1,414 |
| 7 | 276,350 | 274,471 | 1,880 |
| 8 | 294,869 | 292,557 | 2,311 |
| 9 | 311,118 | 308,371 | 2,747 |
| 10 | 325,568 | 322,417 | 3,151 |

*Table 2: The average information content I(x) and I(y) in bits calculated for $N = 100$ e $K = 1$.*

Now to try to understand this result, let's compare the single values $I(x_i)$ and $I(y_i)$ with $f_m(x_i) = y_i$, *|A| = 3*, *K=1* and *N=10*. Therefore, the strings $x \in X^{10}$ have length 10 and having chosen *K=1* the strings $y \in X^{11}$ have length 11. In this situation, the set $X^{10}$ contains $3^{10} = 59049$ strings. We have chosen this value of *N* because it allows us to calculate the single values $I(x_i)$ and $I(y_i)$ and at the same time strings with information content much lower than *NH(X)* contribute negligibly to *I(x)* and *I(y)*.

In Figure 1, the solid line shows the $I(x_i)$ values and the dashed line the $I(y_i)$ values in bits. The strings were sorted according to their information content in ascending order.

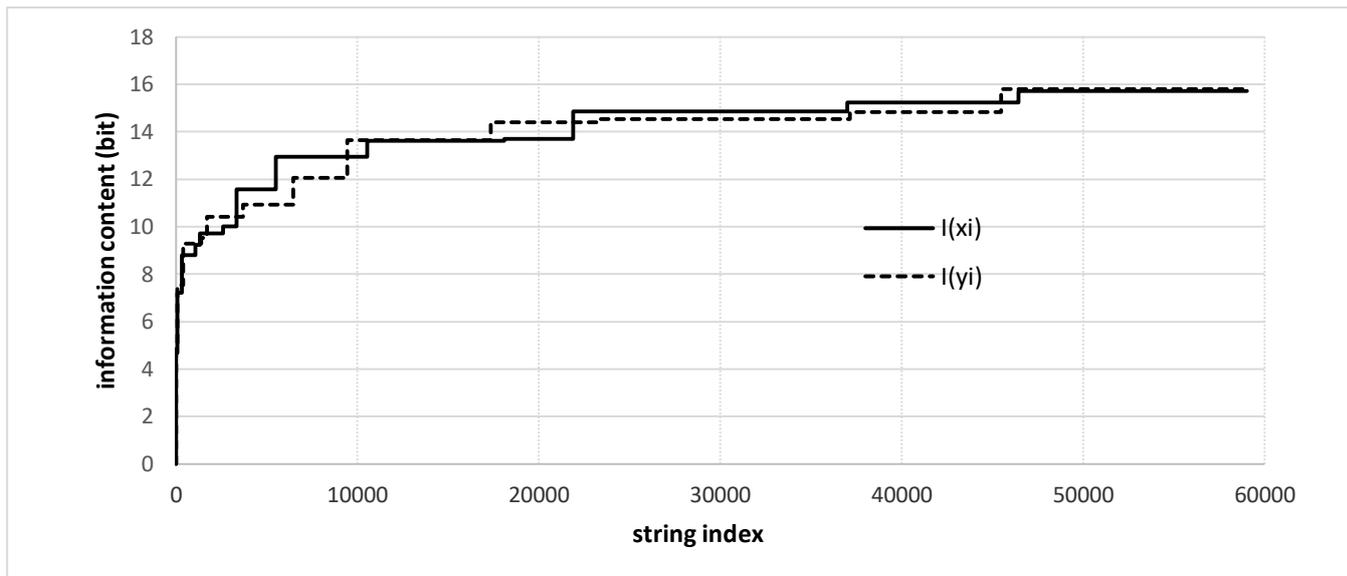

*Figure 1: Comparison between $I(x_i)$ and $I(y_i)$ for N = 100, $|A| = 3$ and K = 1.*

Analyzing Figure 1, we can note that despite $I(y)<I(x)$ ($I(x)=14,263$ bits, $I(y)=14,136$ bits) this inequality is true only on average. Indeed, the single values of $I(x_i)$ and $I(y_i)$ tend to oscillate between them. This result is interesting because it tells us that the use of this technique depends on the probability distribution *P* and consequently on the information content of the typical set. Since the information content of the typical set can be approximated with *NH(X)*, the use of the $f_m$ function can only be useful when this value is placed in an area where $I(y_i) < I(x_i)$.

**Conclusion**

In this article, we have defined the Set Shaping Theory whose goal is the study of the bijection functions that transform a set of strings into a set of equal size made up of strings of greater length. The functions that respect this condition are many but since the goal of this theory is the transmission of data, we have analyzed the function the $f_m$ which transforms the set $X^N$ into the set $Y^{N+k}$ composed of the $|X^N|$ strings with less information content belonging to $X^{N+k}$.

Analyzing the data, we find an unexpected result, indeed the average information content *I(y)* turns out to be less than *I(x)* when the cardinality of *A* is greater than 2. This result is present for minimum lengths such as those reported in table 1 and for longer lengths like the one shown in table 2. Therefore, this result does not seem to depend on the length of the string. However, this is only a preliminary analysis to reach a conclusion it is essential to study the asymptotic behavior.

Figure 1 shows another interesting result, the single values of $I(x_i)$ and $I(y_i)$ oscillate between them and neither of them is greater or less than the other continuously. Consequently, the use of this technique depends on the information content of the typical set.

For these reasons, we believe that this theory is particularly interesting in data compression. However, as mentioned, the consequences of this type of transform on the average information content are particularly complex and therefore this analysis requires further studies.

**Bibliography**

1. C.E. Shannon, "A Mathematical Theory of Communication", *Bell System Technical Journal*, vol. 27, pp. 379–423, 623-656, July, October, 1948.

2. Cover, Thomas M. (2006). *Elements of Information Theory*. John Wiley & Sons. ISBN 0-471-24195-4.

3. Metropolis, N., Ulam, S. (1949). The Monte Carlo Method, Journal of the American Statistical Association. N44 (247). P.335–341. http://www.jstor.org/stable/2280232.

4. Gordon, S.P., & Gordon, F.S. (1990). Applications of Monte Carlo Methods in Calculus. AMATYC Review, v11 n2 P.9-14 Spr 1990. Retrieved from https://eric.ed.gov/?q=EJ411052.